\documentclass[%
preprint,
superscriptaddress,
showpacs,
amsmath,amssymb,
 aps,
 prl,
 floatfix
]{revtex4-1}

\usepackage{graphicx}% Include figure files

\begin{document}

\title{Direct observation of homogeneous cavitation in nanopores }
\author{V.~Doebele}
\affiliation{
 Universit\'e Grenoble Alpes, CNRS, Institut N\'eel, F-38042 Grenoble, France\\
}
\author{A.~Benoit-Gonin}
\affiliation{
 Universit\'e Grenoble Alpes, CNRS, Institut N\'eel, F-38042 Grenoble, France\\
}
\author{F.~Souris}
\affiliation{
 Universit\'e Grenoble Alpes, CNRS, Institut N\'eel, F-38042 Grenoble, France\\
}
\author{L.~Cagnon}
\affiliation{
 Universit\'e Grenoble Alpes, CNRS, Institut N\'eel, F-38042 Grenoble, France\\
}

\author{P.~Spathis}
\affiliation{
 Universit\'e Grenoble Alpes, CNRS, Institut N\'eel, F-38042 Grenoble, France\\
}

\author{P.E.~Wolf}
\email{pierre-etienne.wolf@neel.cnrs.fr}
%, ORCID 0000-0001-8633-7824}
\affiliation{
 Universit\'e Grenoble Alpes, CNRS, Institut N\'eel, F-38042 Grenoble, France\\
}

\author{A.~Grosman}
\email{deceased, 2019 August 29th}
\affiliation{
Sorbonne Universit\'e, CNRS, Institut des NanoSciences de Paris, INSP, F-75005 Paris, France\\
}

\author{M.~Bossert}
\affiliation{
Sorbonne Universit\'e, CNRS, Institut des NanoSciences de Paris, INSP, F-75005 Paris, France\\
}
\author{I.~Trimaille}
\affiliation{
Sorbonne Universit\'e, CNRS, Institut des NanoSciences de Paris, INSP, F-75005 Paris, France\\
}

\author{C.~No\^us}
\affiliation{
Laboratoire Cogitamus 
1 3/4 rue Descartes, 75005 Paris\\
}

\author{E.~Rolley}
\email{rolley@phys.ens.fr}
%, ORCID 0000-0003-1333-2541}
\affiliation{
Laboratoire de Physique de l’Ecole Normale Sup\'erieure, ENS, Universit\'e PSL, CNRS, Sorbonne Universit\'e, Universit\'e de Paris, F-75005 Paris, France\\
}

\date{\today}

\begin{abstract}
We report on the evaporation of hexane from porous alumina and silicon membranes. These membranes contain billions of independent nanopores tailored to an ink-bottle shape, where a cavity several tens of nanometers in diameter is separated from the bulk vapor by a constriction. For alumina membranes with narrow enough constrictions, we demonstrate that cavity evaporation proceeds by cavitation. Measurements of the pressure dependence of the cavitation rate follow the predictions of the bulk, homogeneous, classical nucleation theory, definitively establishing the relevance of homogeneous cavitation as an evaporation mechanism in mesoporous materials. Our results imply that porous alumina membranes are a promising new system to study liquids in a deeply metastable state.
 \end{abstract}

\maketitle

A porous material imbibed with a liquid can dry by two processes: recession of a liquid-vapor interface~\cite{Mason1988a} or formation of vapor bubbles within the material by cavitation~\cite{Ravikovitch2002a, Monson2012a}. Fundamental understanding of which of these processes is effective is crucial for many applications, ranging from characterizing porous materials~\cite{Ravikovitch2002a, Monson2012a} to controlling the shrinkage of concrete~\cite{Maruyama2018b}. 
In particular, this requires establishing whether cavitation occurs in the bulk of pores or on their surface, and, in the first case, whether pores are large enough for homogeneous classical nucleation theory\cite{Blander1975a} (CNT) to hold or whether confinement has to be considered~\cite{Vishnyakov2003a,Vishnyakov2003b,Rasmussen2010a, Bonnet2019a}.

\begin{figure} [!ht]
	\centering
		\includegraphics[width=0.9\textwidth]{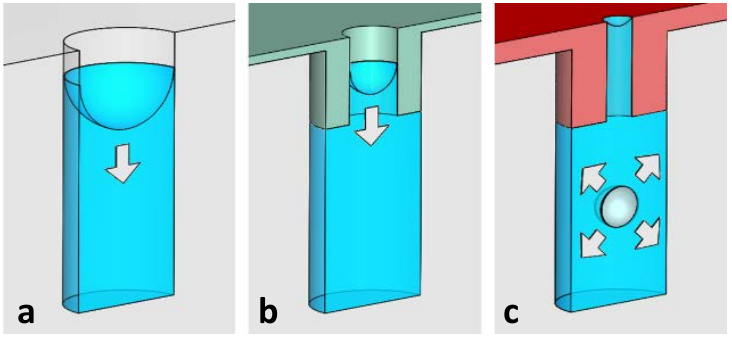}
	\caption{Cavitation in an ink-bottle geometry. (\textbf{a}) Straight cylindrical pore opened to a vapor reservoir, which empties through recession of the liquid-vapor meniscus. 
The smaller the pore diameter, the smaller the evaporation pressure. (\textbf{b},\textbf{c}) Two possible evaporation mechanisms for a cavity ended by a cylindrical constriction. In panel(\textbf{b}), the constriction empties at its equilibrium pressure, triggering further evaporation in the wider cavity through meniscus recession; in panel (\textbf{c}), if the constriction is narrow enough for its evaporation pressure to lie below the cavitation threshold, the cavity empties by cavitation, while the constriction remains filled with liquid.}
	\label{schema}
\end{figure}

Results from previous studies lead to contradictory conclusions. 
As illustrated by Fig.\ref{schema}, evaporation by cavitation is expected for pores presenting an ink-bottle geometry, where a wider cavity is separated from the outside vapor  by a constriction narrow enough for capillarity to block the liquid-vapor interface at the cavitation pressure (Fig.~\ref{schema}(c))~\cite{Ravikovitch2002a, Morishige2003a}. Homogeneous-like cavitation has thus been reported in materials with interconnected pores presenting cavities separated by constrictions, realizing such an ink-bottle geometry. These materials are either ordered (SBA-16 mesoporous silica~\cite{ Rasmussen2010a,Morishige2003a,Morishige2006b}, zeolites~\cite{Maruyama2018a}) or disordered   (cements~\cite{ Maruyama2018a},Vycor~\cite{Morishige2009a, Bonnet2019b}, controlled porous glasses~\cite{ Machin1999a,Reichenbach2011a}). In all cases, the evidence for cavitation is only indirect, relying on the  interpretation of light or x-ray scattering data or, more often, on the observation of a sharp drop in the evaporation isotherm at a given pressure which is then compared to some model, usually CNT. However, in these materials,  the pores have a very small diameter (several nanometers). Attractive interaction with walls should then affect the cavitation threshold,  making  the identification of cavitation through comparison to CNT ambiguous.

In contrast, two experiments performed on nanoporous silicon membranes with ink-bottle pores directly evidenced a two-step shape of the evaporation isotherm of nitrogen around 77~K, consistent with a cavitation mechanism, but at a much \textit{larger} pressure than predicted by the CNT~\cite{Wallacher2004a, Grosman2011a}. This increase is suggestive of heterogeneous cavitation \cite{Grosman2011a}, in strong contrast with the results reported for the more complex geometries above. Moreover, a similar discrepancy has been found for the cavitation of dibromomethane in Vycor~\cite{Mitropoulos2015a}. These results cast doubt on the very principle of using extensions of homogeneous CNT~\cite{Ravikovitch2002a, Bonnet2019a} to predict the cavitation threshold in nanoporous materials.

In this paper, we elucidate this paradoxical situation by studying evaporation of hexane from silicon and alumina membranes with pores' transverse size in the range 20-60~nm, large enough to allow a direct comparison to bulk CNT~\cite{Bonnet2019a}. For both membranes, we directly evidence that cavity evaporation proceeds at a well-defined pressure, consistent with cavitation. For silicon membranes, however, the evaporation pressure strongly depends on the  cavity geometry,  inconsistent with homogeneous cavitation. In contrast, for alumina membranes, we accurately check the activated nature of the evaporation process, and show that it is quantitatively described by bulk CNT. This definitively demonstrates the relevance of homogeneous cavitation as an evaporation mechanism in porous materials.  

\begin{figure} [!h]
\centering
\includegraphics[width=0.9\textwidth]{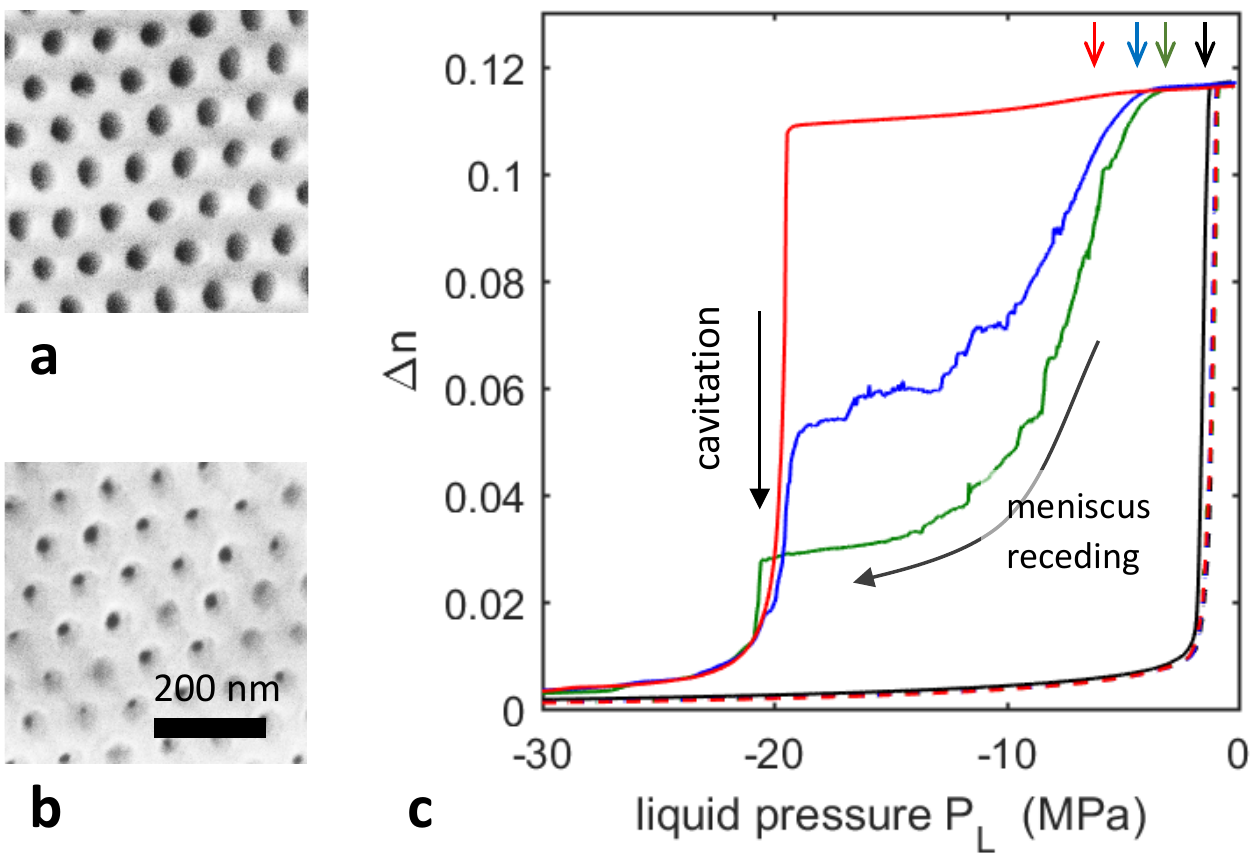}
\caption{Cavitation in an alumina membrane. Starting from a native alumina membrane with 60~nm-diameter pores (\textbf{a}), successive 2-nm-thick alumina layers are deposited at the mouth of the pore, reducing the pore aperture (\textbf{b}: 8 layers). Sorption isotherms of hexane are measured at 18$^{\circ}$C (\textbf{c}) as described in the text. The membrane fluid content is deduced from $\Delta n$, the change of the membrane optical index with respect to the empty state. The (superimposed) dashed curves are the condensation isotherms, and the continuous lines are the evaporation isotherms, for increasing deposits of alumina at the pore mouth (black is for native, green for 8 layers, blue for 10 layers, and red for 12 layers). For intermediate coatings, the noise is due to a loss of contrast of the interference pattern (resulting from a strong light scattering~\cite{SM}). The sharp drop at -\text{20}~MPa is the signature of cavitation.}
\label{isopoAl}
\end{figure}

\begin{figure}[!h]
	\centering
		\includegraphics[width=0.7\textwidth]{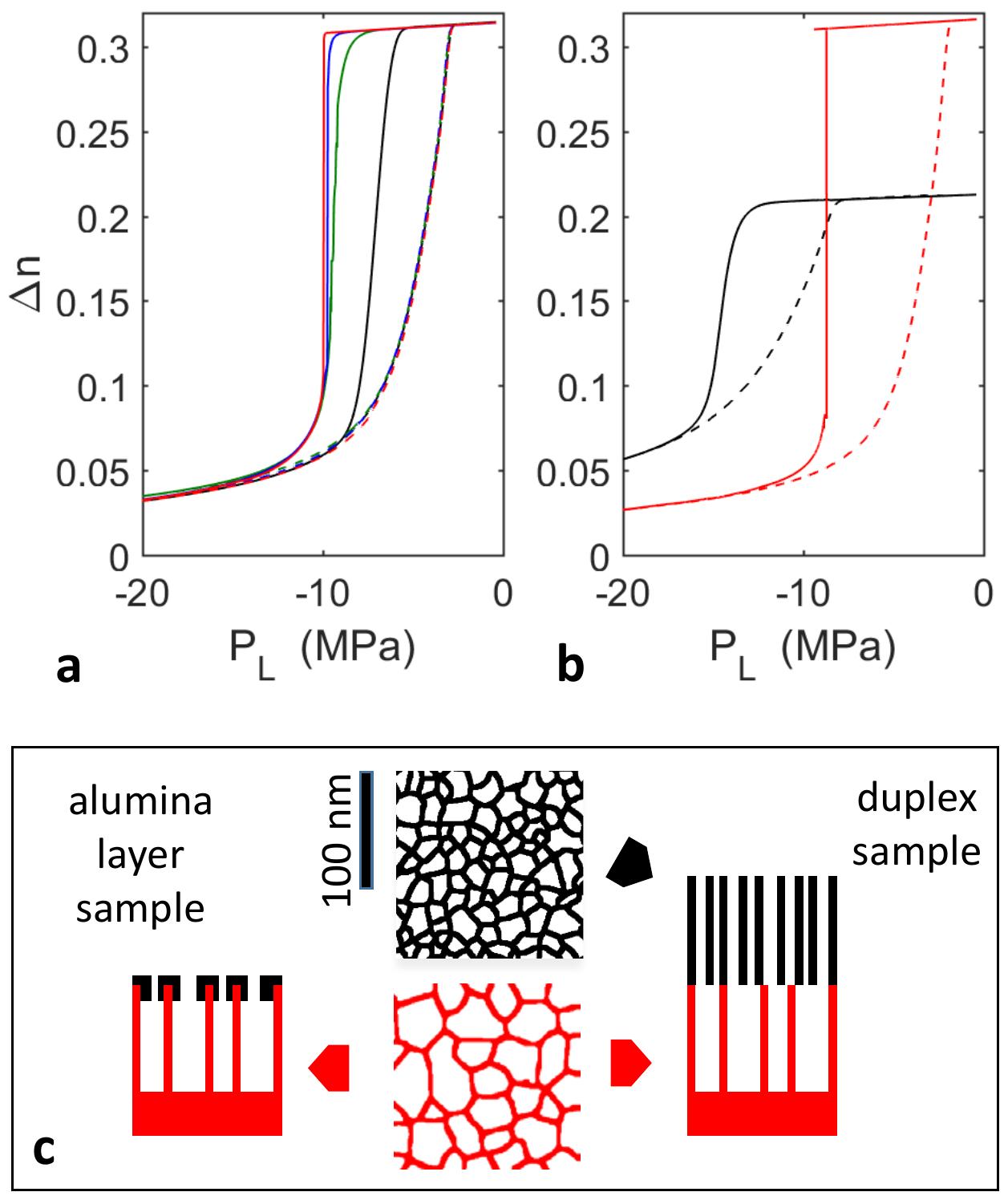}
	\caption{Evaporation in porous silicon ink bottles: (\textbf{a}) Isotherms for an alumina-coated poSi sample (left part of panel \textbf{c}). Superimposed dashed lines correspond to condensation and solid lines to evaporation.  
Black: lines are poSi as prepared ($\langle d \rangle = 26$ nm, pore length $l= 20\, \mathrm{\mu m}$); green/blue/red lines are poSi coated by 2/6/8 alumina layers. For as-prepared poSi, the condensation isotherm is much less steep than in the case of poAl, reflecting the wider distribution of pore diameters. (\textbf{b}) Isotherms obtained on a duplex-layer sample formed by successively electro-etching a top layer with small pores $\langle d \rangle $= 12 nm  and a bottom layer with large pores $\langle d \rangle $= 26 nm (right part  of panel \textbf{c}). The contributions of the two layers are plotted separately in black for the top layer and in red for the bottom layer. (\textbf{c})  binarized transmission electron microscopy (TEM) images of the cross section of cavities  ($\langle d \rangle = 26$ nm, red) and of the constrictions of the duplex sample ($\langle d \rangle = 12$ nm, black.}
	\label{isopoSi}
\end{figure}

Our samples are fabricated using a two step procedure~\cite{SM}. We first synthesize ~$\simeq$1 cm$^2$ nanoporous alumina (poAl) and silicon (poSi) membranes with parallel pores, by electro-etching of highly p-doped silicon~\cite{Grosman2008a} or anodization of aluminum wafers~\cite{Masuda1995a}. In both cases, the pores, about 100 $\mu$m long,  are closed on one side of the membrane and open on the other. As illustrated in Fig.~\ref{isopoAl}(a), the  poAl pores are well organized with a narrow distribution in diameter around an average value of several tens of nanometers, tunable through the anodization conditions~\cite{Lee2008, Lee2014}.  We study two alumina membranes with pore lengths $l=57$ and 76 $\mu$m determined by scanning eectron microscopy (SEM). Their respective average pore diameters are $d=60$~nm and  $d=25$~nm  determined by combining the interpore distance measured by SEM with the membrane porous volume deduced from adsorption isotherms. In contrast, the poSi pores have a polygonal cross section (Fig.~\ref{isopoSi}(c)), with a wider distribution of transverse sizes $d$ around a mean value increasing with the sample porosity~\cite{Grosman2008a} . Most poSi samples have a 70\% porosity corresponding to $d$ in the range 13-40 nm ($\langle d \rangle = 26$ nm) and  $l$ between 5 and 60~$\mu$m. In a second step, we deposit alumina at the pore mouth to obtain the desired ink-bottle geometry. For the 60~nm poAl and poSi samples, we use successive evaporations of 2~nm of aluminum followed by oxidation. For the 25~nm poAl sample,  we used continuous atomic layer deposition (ALD) based on the chemical reaction between trimethylaluminum and water. SEM images show that both methods yield alumina constrictions smaller than 10~nm in diameter \footnote{This is smaller than for previous experiments performed on alumina membranes with ink-bottle pores, reported in Refs.~\cite{Casanova2007a, Bruschi2015a}, explaining why cavitation has not been observed in these experiments.}. This upper bound is consistent with the maximal constriction diameter for observing cavitation, estimated to be 6~nm using Refs.~\citenum{Saam1975a,Israelachvili2011,VanOss1988}~\cite{SM}.

Condensation and evaporation of hexane in these samples were studied in optical cells, regulated at a temperature slightly below the ambient temperature, with a stability of about 1~mK. A capillary line connecting the cell to a tank of hexane at saturated vapor pressure immersed in a temperature-controlled bath is used to fill or empty the membrane  through a precision microvalve at a very small flow rate. The vapor pressure $P_V$ in the cell is measured  by a pressure gauge. The amount of fluid in the pores is determined through the change $\Delta n$ of the membrane optical index measured by white light interferometry (WLI)~\cite{Pacholski2005a,Casanova2007a}.

Figure~\ref{isopoAl} shows the successive sorption isotherms measured for the 60~nm poAl membrane as the pore aperture is progressively reduced. Here, $P_{\rm {V}}$ is converted to $P_{\rm {L}}=(RT/v_{\rm {L}})\ln(P_V/P_{sat})$, the pressure of the liquid in equilibrium with the vapor under the assumptions of  ideal gas and  incompressible liquid ($R$ is the gas constant, $T$ the temperature and $v_{\rm {L}}$ the liquid molar volume).
Condensation takes place at a well-defined pressure, independent of the pore aperture as expected for pores closed at one end. For the native membrane, evaporation occurs at a slightly lower pressure, probably due to some pore corrugation~\cite{Puibasset2007a, Bruschi2015a, Morishige2016a,Doebele_PhD}. Progressively reducing the pore aperture shifts the evaporation to much lower pressures, in agreement with the expectation that evaporation is controlled by meniscus recession in constrictions (Fig.~\ref{schema}(b)). It also broadens the pressure range over which evaporation takes place, showing that the constrictions are distributed in diameter, either due to the initial pore diameter distribution and/or uneven deposition. The salient observation is that, for all coated samples, a sharp drop of the liquid content is observed at the \textit{same} pressure,  $P_{\rm {cav}}\simeq$ -\text{20}~MPa, irrespective of the aperture reduction. The fraction of pores emptying at this pressure increases at each deposition step, and reaches nearly 100\% for the last step. Since, at this stage,  the constrictions necessarily remain distributed in diameter, the fact that the  evaporation pressure is \textit{sharply defined} demonstrates that evaporation takes place within the cavities, with the constrictions remaining filled, in agreement with the cavitation mechanism depicted in Fig.~\ref{schema}(c). We stress that, in contrast to most experiments with porous materials, this evidence is direct, and does not rely on comparing the evaporation pressure value to any model. 

We observe a similar behavior in poSi ink bottles prepared in the same way as poAl (Fig.~\ref{isopoSi}(a)). However, in this case,  the limiting evaporation pressure $P_{\rm {evap}}$ is much larger, around -\text{10}~MPa. 
In order to confirm this difference, 
 we perform complementary experiments on duplex-layer poSi membranes, similar to those previously used for studying nitrogen evaporation~\cite{Grosman2011a}. A bottom layer with large pores --  the cavities --  is connected to the vapor reservoir through a top layer with narrow pores -- the constrictions -- (Fig.~\ref{isopoSi}(c)).  These constrictions are much longer than those obtained by alumina deposition, allowing WLI to \textit{simultaneously} monitor the fluid content in the constrictions and the cavities~\cite{Casanova2007a}.  As shown in Fig.\ref{isopoSi}b, and similarly to Refs.\citenum{Wallacher2004a, Grosman2011a}, cavities empty before the constrictions.  When the length $l$ and diameter $d$ of the cavities are identical to those for the alumina-coated poSi sample ($\langle d \rangle = 26$ nm, $l=20\,\mathrm{\mu m}$), we find $P_{\rm {evap}}\simeq$ --9.5~MPa, close to the --10~MPa obtained in the latter case, showing that this value is intrinsic of the cavity layer.  

On the face of these results, we could be tempted to conclude, as in Ref.\citenum{Grosman2011a}, that evaporation in poSi samples proceeds by heterogeneous cavitation on the cavity surface \cite{elasticity}. However, this is surprising as hexane is believed to perfectly wet most types of surfaces and should nucleate homogeneously. Moreover, as detailed in the Supplemental Material,  $P_{\rm {evap}}$ in poSi can be changed from  --6~MPa to --12~MPa by varying the cavity diameter and length.  This large change seems inconsistent with a heterogeneous cavitation mechanism where the cavitation pressure would only weakly depend on the pore surface. An alternative scenario, suggested by recent NMR studies \cite{Puibasset2016a, Kondrashova2017a}, could be that the cavities of the bottom layer communicate with one another and are connected to the vapor by a small number of wide channels through the top layer (the total volume of these channels being too small to be detected by WLI). In this case, the bottom layer would empty by a percolation mechanism\cite{Mason1988a}, consistent with the observation of a sharply defined evaporation pressure. Testing such a scenario would require further studies.

In contrast, for poAl, repeating the experiment of Fig.\ref{isopoAl} with pores of smaller diameter ($d\simeq 25$ nm) yields nearly the same value $\simeq -20 $ MPa for $P_{\rm{cav}}$ although the pore volume $V_{\rm{p}}$ differs by a factor close to 3 between the two membranes. This observation is consistent with homogeneous cavitation, for which $P_{\rm{cav}}$ depends only weakly on the available volume. It is also consistent with the absence of confinement effect in this diameter range, as  estimated from the model of Ref.\citenum{Bonnet2019a}~\cite{SM}. 

\begin{figure} [!h]
\centering
\includegraphics[width=0.8\textwidth]{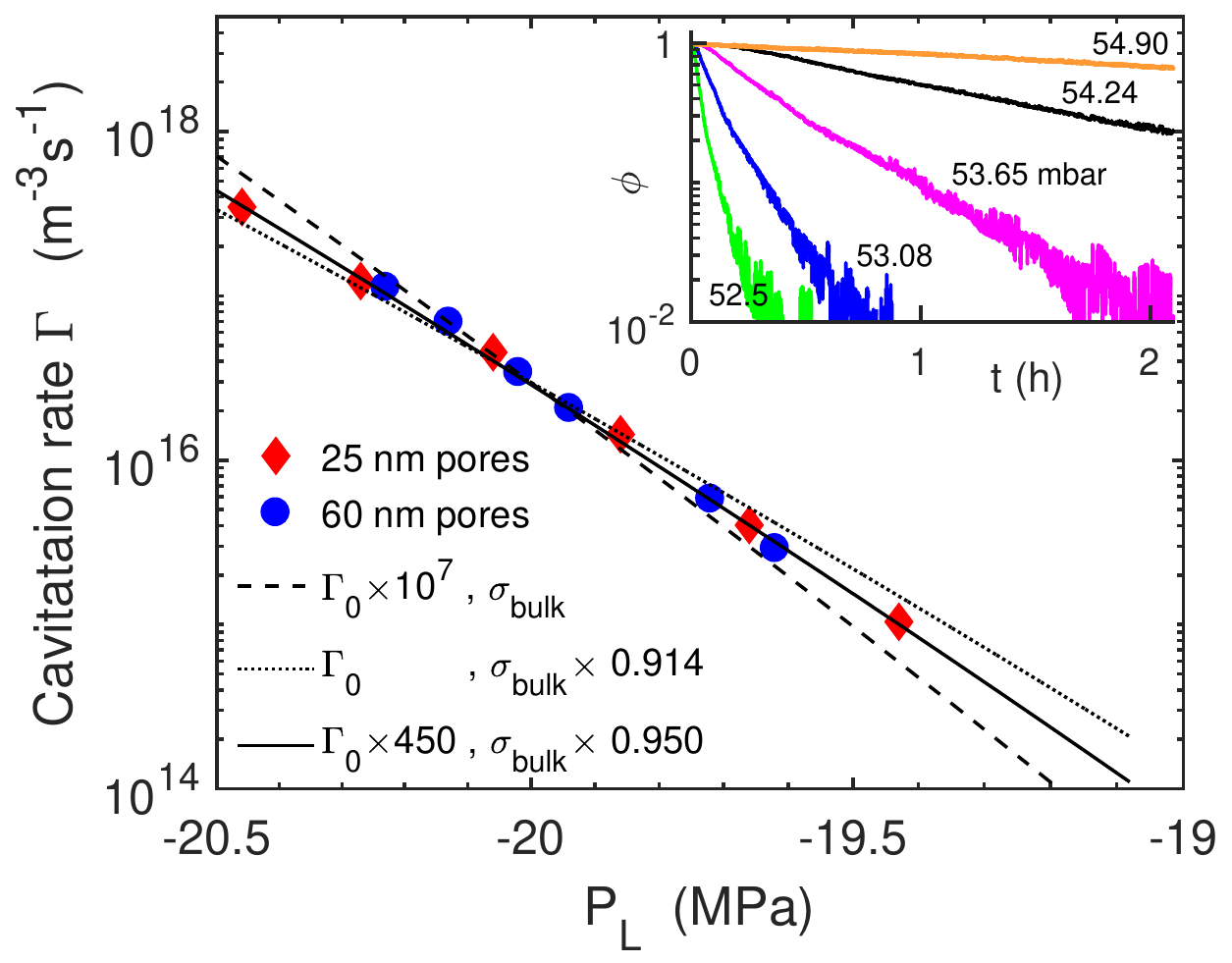}
\caption{Cavitation rate $\Gamma$  as a function of  liquid pressure $P_{L}$ for poAl membranes with average pore diameters $d =60$~nm and $d=25$~nm at 19$^{\circ}$C. Here, $\Gamma$  is measured from the exponential decay of the number of filled pores following a quench of the pressure reservoir from 60~mb down to a lower pressure ranging between 52 and 55 mbar, as illustrated in the inset for the 60-nm membrane. The fraction of filled pores $\phi$ is measured through the logarithm of the optical transmission and normalized to its value at time $t$=0, corresponding to a 10\% transmission(see Supplemental Information). The cavitation rate per unit volume $\Gamma$ is deduced by dividing the decay rate by the pore volume, which is computed using $d=56$~nm and $d=27$~nm. These values lie within the error bars of the measured values above, and are such that $\Gamma (P_L)$ is identical for the two membranes. Lines corresponding to the CNT predictions for different values of the attempt rate or the surface tension (see text). }
\label{tau}
\end{figure}

In order to test whether cavitation in poAl is quantitatively described by CNT, we measure its nucleation rate  $\Gamma (P_{\rm {L}})$. A specific feature of our experimental system is that the membranes contain a very large number of \textit{independent} pores, of the order of several $10^{10}$/cm$^{2}$. 
This feature allows us to  determine the cavitation rate in a single-shot experiment, in contrast to acoustically driven~\cite{Pettersen1994a} or thermally controlled~\cite{Azouzi2013a} bulk cavitation experiments, where the cavitation statistics is determined over thousands of cycles. To this aim, starting from a slightly larger pressure, we quench the reservoir pressure to a stable value corresponding to $P_{\rm{L}}$ around  -\text{20}~MPa, and monitor the temporal decay of the number of filled pores by measuring the light transmission through the membrane. The totally empty or filled membrane, being nearly homogeneous on the scale of the light wavelength, only weakly  scatters. In contrast, when stochastic cavitation takes place,  the pores randomly empty, giving rise to local fluctuations of the refractive index and strong light scattering, resulting in a reduced transmission. This effect can be directly evidenced by illuminating the membrane with a wide collimated light beam and precisely quantified by measuring the transmission of a laser beam~\cite{SM}.

We have performed such experiments for the two alumina membranes. As expected for a stochastic process, the number of filled pores decreases exponentially with time with a time constant $\tau$ (inset of Fig.~\ref{tau} for the 60-nm membrane). Repeating this quench at different depths yields $\tau(P_{\rm{L}})$, from which we deduce the cavitation rate per unit time and unit volume $\Gamma(P_{\rm{L}})=( V_{\rm{p}}\,\tau(P_{\rm{L}}))^{-1}$.
 
Figure~\ref{tau} shows that  $\Gamma(P_{\rm{L}})$ increases by nearly 3 orders of magnitude when changing $P_{\rm{V}}$, hence $P_{\rm{L}}$, by only 5\%. This large increase exemplifies the exponential dependence of the relaxation time on the energy barrier. 

To our knowledge, these results are the first evidence for the stochastic and activated nature of cavitation in nanoporous materials. The relaxation rate can be compared to the CNT prediction~\cite{Blander1975a} $\Gamma=\Gamma_0 \exp (-E_{\rm{b}}/k_{\rm B}T)$, where $\Gamma_0$ is an attempt rate and the energy barrier $E_{\rm{b}}$ is given by:

\begin {equation}
E_{\rm{b}}=\frac{16\pi\sigma^3}{3(P_{\rm{V}}-P_{\rm{L}})^2}
\label{eqEb}
\end{equation}
where $\sigma$ is the surface tension. Different expressions for $\Gamma_0$~\cite{Blander1975a, Maris1989a} lead to $\Gamma_0$ = 2.10$^{38}$~m$^{-3}$~s$^{-1}$ within a factor of 10. As shown by Fig.\ref{tau}, using this value and the bulk surface tension (0.185 J/m$^2$ at 19$^\circ$C~\cite{NIST}) leads to predicted rates that are about 10$^7$ too small over the full pressure range, corresponding to a predicted cavitation pressure 15\% larger than observed. Keeping $\Gamma_0$ = 2.10$^{38}$~m$^{-3}$~s$^{-1}$, an approximate agreement with experiments requires us to reduce the surface tension $\sigma$ by about 9\% with respect to its bulk value. A similar difference has been measured in the case of bulk cavitation of heptane and ascribed to a reduction of the surface tension due to the large curvature of the critical germ~\cite{Bruot2016a}. However, adjusting only $\Gamma_0$ or $\sigma$ does not allow us to match the experimental  $\Gamma$ vs $P_{\rm{L}}$ data over the full pressure range. Such a match requires that we combine a 5\% reduction of $\sigma$ \textit{with} an increase of $\Gamma_0$ by a factor of order 500. This requirement might point to the fact that the above expressions of the CNT attempt rate are underestimated.

To summarize, we show that cavitation of hexane in porous alumina ink bottles with cavity diameters of several tens of nanometers closely follows the predictions of the bulk, homogeneous, classical nucleation theory. This result confirms that homogeneous cavitation is a relevant mechanism  in nanoporous systems, as assumed by previous studies on materials with interconnected pores. Our approach opens the way to performing similar experiments with pores of smaller diameter (below 10~nm) in order to quantitatively test extensions of CNT in the presence of confinement. In contrast, we confirm earlier experiments,  finding that porous silicon ink bottles empty more easily than expected from CNT . More studies will be required to understand whether this is due to heterogeneous cavitation or to a nonideal structure of poSi.  
   
Finally, our work opens new prospects for fundamental studies of cavitation. In contrast to cavitation acoustically driven at MHz frequencies, our experiments are essentially static, allowing us  to precisely measure the relaxation rate at a given pressure. Also, in contrast to the so-called artificial-tree technique~\cite{Wheeler2008a}, where cavitation is probed in macroscopic cavities closed by a single porous layer, nanoporous membranes with independent nanopores tolerate the existence of a small density of leaky constrictions. PoAl membranes are thus a promising system to address points of current debate, such as the influence of superfluidity of liquid helium on its cavitation~\cite{Qu2015a}, or the origin of the much larger cavitation pressure observed for water in the artificial-tree geometry~\cite{Caupin2012a} as compared to that observed in quartz inclusions~\cite{Zheng1991a}. Beyond these examples, by allowing us to decrease the liquid pressure down to its tensile limit, these membranes open a new route to study liquids in deeply metastable states.

This paper is dedicated to the memory of our late co-author and colleague A.~Grosman. We thank K. Davitt for her critical reading. We acknowledge the financial support of Agence Nationale de la Recherche through the project CavConf, ANR-17-CE30-0002, including the funding of F. S. and M. B., and of Universit\'e Grenoble Alpes, which funded V.D. through an AGIR PhD grant. A. G., M. B., I. T., E. R., V. D., A. B.-G., F. S., P. S., and P.~E.~W.participated in the measurements and analyses; P. E. W. and E. R. wrote the paper with contributions from L. C., and P. S. C. N. represents a consortium who has contributed to the conceptualization and methodology of this work.


\begin{thebibliography}{100}

\bibitem{Mason1988a}
Mason G.
\newblock Determination of the pore-size distributions and pore-space
  interconnectivity of Vycor porous glass from adsorption-desorption hysteresis
  capillary condensation isotherms.
\newblock \emph{Proc. R. Soc. A}, \textbf{415}, 453 (1988).

\bibitem{Ravikovitch2002a}
Ravikovitch P and Neimark A.
\newblock {Experimental confirmation of different mechanisms of evaporation
  from ink-bottle type pores: Equilibrium, pore blocking, and cavitation}.
\newblock \emph{Langmuir}, \textbf{18}, 9830 (2002).

\bibitem{Monson2012a}
Monson P.
\newblock Understanding adsorption/desorption hysteresis for fluids in
  mesoporous materials using simple molecular models and classical density
  functional theory.
\newblock \emph{Micropor. Mesopor. Mat.}, \textbf{160}, 47  (2012).

\bibitem{Maruyama2018b}
Maruyama I, Gartner E, Beppu K, and Kurihara R.
\newblock Role of alcohol-ethylene oxide polymers on the reduction of shrinkage
  of cement paste.
\newblock \emph{Cement and Concrete Research}, \textbf{111}, 157  (2018).

\bibitem{Blander1975a}
Blander M and Katz J.
\newblock {Bubble nucleation in liquids}.
\newblock \emph{AIChE Journal}, \textbf{21}, 833 (1975).

\bibitem{Vishnyakov2003a}
Vishnyakov A and Neimark A.
\newblock Monte Carlo simulation test of pore blocking effects.
\newblock \emph{Langmuir}, \textbf{19}, 3240 (2003).

\bibitem{Vishnyakov2003b}
Vishnyakov A and Neimark A.
\newblock Nucleation of liquid bridges and bubbles in nanoscale capillaries.
\newblock \emph{J. Chem. Phys.}, \textbf{119}, 9755 (2003).

\bibitem{Rasmussen2010a}
Rasmussen C~J, Vishnyakov A, Thommes M, Smarsly B~M, Kleitz F, and Neimark A~V.
\newblock {Cavitation in Metastable Liquid Nitrogen Confined to Nanoscale
  Pores}.
\newblock \emph{Langmuir}, \textbf{26}, 10147 (2010).

\bibitem{Bonnet2019a}
Bonnet F and Wolf P~E.
\newblock Thermally activated condensation and evaporation in cylindrical
  pores.
\newblock \emph{J. Phys. Chem. C}, \textbf{123}, 1335
  (2019).

\bibitem{Morishige2003a}
Morishige K and Tateishi N.
\newblock {Adsorption hysteresis in ink-bottle pore}.
\newblock \emph{J. Chem. Phys.}, \textbf{119}, 2301 (2003).

\bibitem{Morishige2006b}
Morishige K, Tateishi M, Hirose F, and Aramaki K.
\newblock Change in desorption mechanism from pore blocking to cavitation with
  temperature for nitrogen in ordered silica with cagelike pores.
\newblock \emph{Langmuir}, \textbf{22}, 9220 (2006).

\bibitem{Maruyama2018a}
Maruyama I, Ryme{\v{s}} J, Vandamme M, and Coasne B.
\newblock Cavitation of water in hardened cement paste under short-term
  desorption measurements.
\newblock \emph{Materials and Structures}, \textbf{51}, 159 (2018).

\bibitem{Morishige2009a}
Morishige K.
\newblock {Hysteresis Critical Point of Nitrogen in Porous Glass: Occurrence of
  Sample Spanning Transition in Capillary Condensation}.
\newblock \emph{Langmuir}, \textbf{25}, 6221 (2009).

\bibitem{Bonnet2019b}
Bonnet F, Melich M, Puech L, Angl\`{e}s~d'Auriac J~C, and Wolf P~E.
\newblock On condensation and evaporation mechanisms in disordered porous
  materials.
\newblock \emph{Langmuir}, \textbf{35}, 5140 (2019).

\bibitem{Machin1999a}
Machin W~D.
\newblock Properties of three capillary fluids in critical region.
\newblock \emph{Langmuir}, \textbf{15}, 169 (1999).

\bibitem{Reichenbach2011a}
Reichenbach C, Kalies G, Enke D, and Klank D.
\newblock Cavitation and pore blocking in nanoporous glasses.
\newblock \emph{Langmuir}, \textbf{27}, 10699 (2011).

\bibitem{Wallacher2004a}
Wallacher D, Kunzner N, Kovalev D, Knorr N, and Knorr K.
\newblock Capillary condensation in linear mesopores of different shape.
\newblock \emph{Phys. Rev. Lett.}, \textbf{92}, 195704 (2004).

\bibitem{Grosman2011a}
Grosman A and Ortega C.
\newblock {Cavitation in Metastable Fluids Confined to Linear Mesopores}.
\newblock \emph{Langmuir}, \textbf{{27}}, 2364 (2011).

\bibitem{Mitropoulos2015a}
Mitropoulos A~C, Stefanopoulos K~L, Favvas E~P, Vansant E, and Hankins N~P.
\newblock On the formation of nanobubbles in Vycor porous glass during the
  desorption of halogenated hydrocarbons.
\newblock \emph{Scientific Reports}, \textbf{5}, 10943 (2015).

\bibitem{SM}
See Supplemental Material for  details on the sample preparation, the measurements methods, and the analysis of the results. 

\bibitem{Grosman2008a}
Grosman A and Ortega C.
\newblock Capillary condensation in porous materials. hysteresis and
  interaction mechanism without pore blocking/percolation process.
\newblock \emph{Langmuir}, \textbf{24}, 3977 (2008).

\bibitem{Masuda1995a}
Masuda H and Fukuda K.
\newblock Ordered metal nanohole arrays made by a two-step replication of
  honeycomb structures of anodic alumina.
\newblock \emph{Science}, \textbf{268}, 1466 (1995).

\bibitem{Lee2008}
Lee W, Schwirn K, Steinhart M, Pippel E, Scholz R, and Gosele U.
\newblock Structural engineering of nanoporous anodic aluminium oxide by pulse
  anodization of aluminium.
\newblock \emph{Nat. Nano}, \textbf{3}, 234 (2008).

\bibitem{Lee2014}
Lee W and Park S~J.
\newblock Porous anodic aluminum oxide: Anodization and templated synthesis of
  functional nanostructures.
\newblock \emph{Chem. Rev.}, \textbf{114}, 7487 (2014).


\bibitem{Note1}
This is smaller than for previous experiments performed on alumina membranes
  with ink-bottle pores, reported in Refs.\cite {Casanova2007a, Bruschi2015a},
  explaining why cavitation has not been observed in these experiments.

 \bibitem{Casanova2007a}
Casanova F, Chiang C~E, Li C~P, and Schuller I~K.
\newblock Direct observation of cooperative effects in capillary condensation:
  The hysteretic origin.
\newblock \emph{Appl. Phys. Lett.}, \textbf{91}, 243103 (2007).

\bibitem{Bruschi2015a}
Bruschi L, Mistura G, Nguyen P~T~M, Do D~D, Nicholson D, Park S~J, and Lee W.
\newblock Adsorption in alumina pores open at one and at both ends.
\newblock \emph{Nanoscale}, \textbf{7}, 2587 (2015).


\bibitem{Saam1975a}
Saam W~F and Cole M~W.
\newblock Excitations and thermodynamics for liquid-helium films.
\newblock \emph{Phys. Rev. B}, \textbf{11}, 1086 (1975).

\bibitem{Israelachvili2011}
Israelachvili J~N.
\newblock Intermolecular and Surface Forces (Third Edition),  (Academic Press,
  San Diego 2011).


\bibitem{VanOss1988}
Van~Oss C~J, Chaudhury M~K, and Good R~J.
\newblock Interfacial Lifshitz-Van der Waals and polar interactions in
  macroscopic systems.
\newblock \emph{Chem. Rev.}, \textbf{88}, 927 (1988).


\bibitem{Pacholski2005a}
Pacholski C, Sartor M, Sailor M~J, Cunin F, and Miskelly G~M.
\newblock Biosensing using porous silicon double-layer interferometers:
  Reflective interferometric Fourier transform spectroscopy.
\newblock \emph{J. Am. Chem. Soc.}, \textbf{127}, 11636
  (2005).
  

\bibitem{Puibasset2007a}
Puibasset J.
\newblock Adsorption/desorption hysteresis of simple fluids confined in
  realistic heterogeneous silica mesopores of micrometric length: A new
  analysis exploiting a multiscale Monte Carlo approach.
\newblock \emph{J. Chem. Phys.}, \textbf{127}, 154701 (2007).


\bibitem{Morishige2016a}
Morishige K.
\newblock Nature of adsorption hysteresis in cylindrical pores: Effect of pore
  corrugation.
\newblock \emph{J. Phys. Chem. C}, \textbf{120}, 22508
  (2016).

\bibitem{Doebele_PhD}
Doebele V.
\newblock Condensation et \'{e}vaporation de l'hexane dans les membranes
  d'alumine poreuse.
\newblock Ph.D. thesis, Universit{\'e} Grenoble-Alpes (2019).

\bibitem{elasticity}
As discussed in the Supplemental Material, detailed measurements of the stress-induced deformation of poSi \cite{Grosman2015a, Rolley2017a,Bossert2020a} show that it is too small to account for the measured evaporation pressure.

\bibitem{Grosman2015a}
Grosman A, Puibasset J, and Rolley E.
\newblock Adsorption-induced strain of a nanoscale silicon honeycomb.
\newblock \emph{Europhys. Lett.}, \textbf{109}, 56002 (2015).

\bibitem{Rolley2017a}
Rolley E, Garroum N, and Grosman A.
\newblock Using capillary forces to determine the elastic properties of
  mesoporous materials.
\newblock \emph{Phys. Rev. B}, \textbf{95}, 064106 (2017).


\bibitem{Bossert2020a}
Bossert M, Grosman A, Trimaille I, No{\^u}s C, and Rolley E.
\newblock Stress or strain does not impact sorption in stiff mesoporous
  materials.
\newblock \emph{Langmuir}, \textbf{36}, 11054 (2020).
%\newblock PMID: 32841029.

\bibitem{Puibasset2016a}
{Puibasset, J}, {Porion, P}, {Grosman, A}, and {Rolley, E}.
\newblock Structure and permeability of porous silicon investigated by
  self-diffusion nmr measurements of ethanol and heptane.
\newblock \emph{Oil Gas Sci. Technol. - Rev. IFP Energies nouvelles},
  \textbf{71}, 54 (2016).

\bibitem{Kondrashova2017a}
Kondrashova D, Lauerer A, Mehlhorn D, Jobic H, Feldhoff A, Thommes M,
  Chakraborty D, Gommes C, Zecevic J, de~Jongh P, Bunde A, K{\"a}rger J, and
  Valiullin R.
\newblock Scale-dependent diffusion anisotropy in nanoporous silicon.
\newblock \emph{Scientific Reports}, \textbf{7}, 40207 (2017).


\bibitem{Pettersen1994a}
Pettersen M~S, Balibar S, and Maris H~J.
\newblock Experimental investigation of cavitation in superfluid
  $^{4}\mathrm{He}$.
\newblock \emph{Phys. Rev. B}, \textbf{49}, 12062 (1994).

\bibitem{Azouzi2013a}
Azouzi M~E~M, Ramboz C, Lenain J~F, and Caupin F.
\newblock A coherent picture of water at extreme negative pressure.
\newblock \emph{Nat. Phys.}, \textbf{9}, 38 (2013).

\bibitem{Maris1989a}
Maris H~J and Xiong Q.
\newblock Nucleation of bubbles in liquid helium at negative pressure.
\newblock \emph{Phys. Rev. Lett.}, \textbf{63}, 1078 (1989).

\bibitem{NIST}
Lemmon E. W., McLinden M. O. and Friend D. G., "Thermophysical Properties of Fluid Systems" in WebBook of Chemistry  NIST, NIST Standard Reference Database SRD Number 69, Eds. P.J. Linstrom and W.G. Mallard, National Institute of Standards and Technology, Gaithersburg MD, 20899, https://doi.org/10.18434/T4D303, (downloaded April 10th, 2020).

\bibitem{Bruot2016a}
Bruot N and Caupin F.
\newblock Curvature dependence of the liquid-vapor surface tension beyond the Tolman approximation.
\newblock \emph{Phys. Rev. Lett.}, \textbf{116}, 056102 (2016).

\bibitem{Wheeler2008a}
Wheeler T~D and Stroock A~D.
\newblock The transpiration of water at negative pressures in a synthetic tree.
\newblock \emph{Nature}, \textbf{455}, 208 (2008).

\bibitem{Qu2015a}
Qu A, Trimeche A, Dupont-Roc J, Grucker J, and Jacquier P.
\newblock Cavitation density of superfluid helium-4 around 1 K.
\newblock \emph{Phys. Rev. B}, \textbf{91}, 214115 (2015).

\bibitem{Caupin2012a}
Caupin F, Arvengas A, Davitt K, Azouzi M~E~M, Shmulovich K~I, Ramboz C, Sessoms D~A, and Stroock A~D.
\newblock Exploring water and other liquids at negative pressure.
\newblock \emph{J. Phys. Cond. Mat.}, \textbf{24}, 284110 (2012).

\bibitem{Zheng1991a}
Zheng Q, Durben D~J, Wolf G~H, and Angell C~A.
\newblock Liquids at large negative pressures: Water at the homogeneous
  nucleation limit.
\newblock \emph{Science}, \textbf{254}, 829 (1991).

\end{thebibliography}
\end{document}